\documentclass[manuscript]{emulateapj}
\usepackage{graphicx}
\usepackage{natbib}
\usepackage{subfigure}

\begin{document}

\title{Probing Pre-galactic Metal Enrichment with High-Redshift Gamma-Ray Bursts}

\author{F. Y. Wang$^{1,2,7}$, Volker Bromm$^{2,3}$, Thomas H. Greif$^4$, Athena Stacy$^5$, Z. G. Dai$^1$, Abraham Loeb$^6$ and K. S. Cheng$^7$}

\affil{$^1$School of Astronomy and Space Science, Nanjing
University, Nanjing 210093, China
\\ $^2$Department of Astronomy, University of Texas at Austin,
Austin, TX 78712, USA \\ $^3$Texas Cosmology Center, University of
Texas at Austin, TX 78712, USA \\ $^4$Max-Planck-Institut f\"{u}r
Astrophysik, Karl-Schwarzschild-Strasse 1, 85740 Garching bei
M\"{u}nchen, Germany\\ $^5$NASA Goddard Space Flight Center, Greenbelt,
MD 20771, USA
\\ $^6$Astronomy Department, Harvard University, 60 Garden St., Cambridge,
MA 02138, USA \\ $^7$Department of Physics, The University of Hong
Kong, Pokfulam Road, Hong Kong, China}

\begin{abstract}

We explore high-redshift gamma-ray bursts (GRBs) as promising tools
to probe pre-galactic metal enrichment. We utilize the bright
afterglow of a Population~III (Pop III) GRB exploding in a
primordial dwarf galaxy as a luminous background source, and calculate
the strength of metal absorption lines that are imprinted by the
first heavy elements in the intergalactic medium (IGM). To derive the
GRB absorption line diagnostics, we use an existing highly-resolved
simulation of the formation of a first galaxy which is characterized
by the onset of atomic hydrogen cooling in a halo with
virial temperature $\ga 10^4$\,K. We explore the unusual
circumburst environment inside the systems that hosted Pop~III stars,
modeling the density evolution with the self-similar solution for a
champagne flow. For minihalos close to the cooling threshold, the
circumburst density is roughly proportional to $(1+z)$
with values of about a few $\rm cm^{-3}$. In more massive halos, corresponding
to the first galaxies, the density may be larger,
$n\ga 100$\,cm$^{-3}$. The resulting afterglow fluxes are weakly dependent on
redshift at a fixed observed time, and may be detectable with the
\emph{James Webb Space Telescope (JWST)} and Very Large Array (VLA)
in the near-IR and radio wavebands, respectively, out to redshift $z\ga 20$.
We predict that the maximum of the afterglow emission shifts
from near-IR to millimeter bands with peak
fluxes from mJy to Jy at different observed times. The metal absorption
line signature is expected to be detectable in the near future.
GRBs are ideal
tools for probing the metal enrichment in the early IGM, due to
their high luminosities and featureless power-law spectra. The metals in
the first galaxies produced by the first supernova (SN) explosions are
likely to reside in low-ionization stages (\ion{C}{2}, \ion{O}{1},
\ion{Si}{2} and \ion{Fe}{2}). We show that if the afterglow can be observed
sufficiently early, analysis of the metal lines may distinguish
whether the first heavy elements were produced in a
pair-instability supernova (PISN), or a core-collapse (Type~II) SN, thus
constraining the initial mass function of the first stars.

\end{abstract}


\section{Introduction}
One of the key goals in modern cosmology is to study the formation
of the first stars and galaxies at the
end of the cosmic dark ages, and how they shaped the subsequent evolution
of the universe (Barkana \& Loeb 2001;
Bromm \& Larson 2004; Ciardi \& Ferrara 2005; Bromm et al. 2009;
Loeb 2010). The first, so-called Population~III (Pop~III) stars
are predicted to have formed at $z\ga 20$ in minihalos with virial mass
$M_{\rm vir} \sim 10^6 M_{\odot}$ and temperatures
$T_{\rm vir} < 10^4$\,K (Haiman et al. 1996;
Tegmark et al. 1997; Yoshida et al. 2003). The Pop~III initial mass
function (IMF) is thought to be top-heavy (Bromm et al. 1999, 2002;
Abel et al. 2002), possibly extending to
$M_*\ga 100 M_{\sun}$, but recent simulations indicate that the Pop~III
IMF may be quite broad, also including a fraction of lower-mass
stars (Stacy et al. 2010; Clark et al. 2011b; Greif et al. 2011, 2012). The first bona-fide galaxies
are expected to have formed at a later stage in hierarchical structure
formation (Bromm \& Yoshida 2011), when
$\sim 10^8 M_{\sun}$ halos assembled at $z\ga 10$ via the merging
of progenitor minihalos (Wise \& Abel 2007; Greif et al. 2008).
These systems are often termed
`atomic cooling halos', because their virial temperature, $T_{\rm vir}\ga
10^4$\,K, exceeds the threshold to enable efficient cooling via lines
of atomic hydrogen (Oh \& Haiman 2002).
Direct observations of the first galaxies at redshifts $z>10$
have so far been out of reach. In the coming decade, the \emph{James
Webb Space Telescope (JWST)} promises to directly probe this critical
period (Gardner et al. 2006). The detection of metal absorption lines in the
afterglow spectrum of high-redshift gamma-ray bursts (GRBs), imprinted by
enriched gas in the first galaxies, offers an unusual opportunity to
study the physical conditions inside them. We may thus be able to
derive constraints on the
temperature, metallicity, ionization state, and kinematics
in the interstellar medium (ISM) of high-redshift galaxies, and
in the surrounding intergalactic medium (IGM). It is encouraging that
such diagnostics can already be obtained for bursts at somewhat lower
redshifts. An example is GRB~081008, where high-resolution spectroscopy with
the VLT has probed the ISM of a host galaxy at $z\simeq 2$ (D'Elia
et al. 2011).

Long-duration GRBs have been shown to be associated with the death
of massive stars (Stanek et al. 2003; Hjorth et al. 2003; Woosley \&
Bloom 2006). Their high luminosities make them detectable out to the
edge of the visible universe (Ciardi \& Loeb 2000; Lamb \& Reichart
2000; Bromm \& Loeb 2002, 2006; Gou et al. 2004; Inoue et al. 2007;
M\'{e}sz\'{a}ros \& Rees 2010), with the current record held by
GRB~090429B at $z\sim9.4$ (Cucchiara et al. 2011). GRBs provide
ideal probes of the high-redshift universe, including the star
formation rate (Totani 1997; Wijers et al. 1998; Porciani \& Madau
2001; Chary et al. 2007; Y\"{u}ksel et al. 2008; Wang \& Dai 2009, 2011; Elliott
et al. 2012), reionization (Gallerani et al. 2008), dark energy (Dai et al. 2004; Wang et al. 2011),
and the IGM metal enrichment (Barkana \& Loeb 2004; Totani et al.
2006; Toma et al. 2011; Bromm \& Loeb 2012). The leading contender
for the central engine of long-duration GRBs is the collapsar model
(Woosley 1993; MacFadyen et al. 2001). Because of their predicted
high characteristic mass, a significant fraction of Pop~III stars
might end their lives as a black hole, potentially leading to a
large number of high-redshift GRBs. Thus, Pop~III stars are viable
progenitors of long-duration GRBs, triggered by the collapsar
mechanism, as long as they can lose their outer envelope and retain
sufficient angular momentum in their center (Bromm \& Loeb 2006;
Belczynski et al. 2007; Komissarov \& Barkov 2010; Stacy et al. 2011).
It might even be possible for Pop~III collapsars to occur if the extended
outer envelope were not lost (Suwa \& Ioka 2011).

The history of pre-galactic metal enrichment has several important consequences
for structure formation (Madau et al. 2001; Karlsson et al. 2012).
An early phase of metal injection may qualitatively change
the character of star formation, from a predominantly high-mass (Pop~III) mode
to a normal, low-mass dominated (Pop I/II) one, once the enrichment
has exceeded a `critical metallicity' of $Z_{\rm crit}\sim 10^{-4} Z_{\sun}$
(Bromm et al. 2001a; Schneider et al. 2002, 2006; Bromm \& Loeb 2003; Mackey et al.
2003). The transition between these two modes
has crucial implications, e.g., for the expected
redshift distribution of GRBs (Bromm \& Loeb 2006; Campisi et al. 2011; de Souza et al. 2011), for reionization
(Cen 2003; Wyithe \& Loeb 2003; Furlanetto \& Loeb 2005), and for the chemical abundance
patterns of low-metallicity stars (Qian \& Wasserburg 2001; Frebel et al.
2007, 2009; Tumlinson 2010). It is therefore important to explore the topology of early metal
enrichment, and to determine when particular regions in the universe become
supercritical (Tornatore et al. 2007; Maio et al. 2010).

Absorption lines
imprinted on the spectra of bright background sources, such as GRBs
or quasars, are one of the main sources of information about
the physical and chemical properties of high-redshift systems
(Oh 2002; Furlanetto \& Loeb 2003; Oppenheimer et al. 2009).
These lines are due
mainly to absorption by neutral hydrogen present in the low
column-density Ly$\alpha$ forest, and by metals in
low-ionization stages (e.g., \ion{C}{2},
\ion{Si}{2}, \ion{Mg}{2}, \ion{Fe}{2}, \ion{O}{1}) which
arise in the higher column-density gas associated with
Damped Ly$\alpha$ Absorbers (DLAs). The analysis of the
spectrum of distant GRB~050904 (Totani et al. 2006) has resulted
in a wealth of detailed insight into the physical conditions
within the host galaxy at $z\simeq 6.3$, and Salvaterra et al. (2009)
claimed that they identified two
absorption lines (\ion{Si}{4} and \ion{Fe}{2}), although at poor signal-to-noise, in the spectrum
of GRB~090423, the most distant spectroscopically confirmed burst at $z=8.2$ (Salvaterra et al. 2009;
Tanvir et al. 2009).
GRBs as background sources offer a number of advantages compared to
traditional lighthouses such as quasars (Bromm \& Loeb 2012). Their number
density drops much less precipitously than quasars at $z>6$ (Fan et al. 2006),
and the absence of a strong proximity effect, together with the near
power-law character of their spectra, renders them ideal probes of
the early IGM.

In this paper, we discuss the observational signatures of Pop~III
GRBs and study pre-galactic metal enrichment utilizing
absorption lines in the spectra of high-$z$ GRBs which were imprinted
by the first galaxies. Recently, it has become feasible to study the
formation of the first galaxies, including the metal enrichment from
Pop~III supernovae (SNe), with highly-resolved cosmological simulations
(Wise \& Abel 2008; Greif et al. 2010). We here place a bright GRB
into the simulation box of Greif et al. (2010), and derive the spectral
signature as the afterglow light escapes from the first galaxy,
thereby probing the partially enriched IGM in its vicinity.
The fluxes in the near-IR and radio bands may be detectable
by the {\it JWST} and the Very Large Array (VLA) out to
$z\ga 20$.
The structure of this paper is as follows. In Section~2, we derive
the circumburst density profile of Pop~III GRBs, followed by
a brief description of the underlying first galaxy simulation (Section~3).
We discuss the properties of the GRB afterglow in Section~4, and the metal
absorption line diagnostics in Section~5, followed by our conclusions.

\section{GRB density profile}

Two different environments are currently discussed as Pop~III star
formation sites, minihalos and atomic cooling halos (Bromm et al. 2009).
The latter mode is often termed Pop~III.2 to indicate
physical conditions, such as a higher degree of ionization, that differs
from the canonical minihalo case, possibly resulting in somewhat lower
masses (McKee \& Tan 2008).
For normal GRBs, either a
constant number-density profile, or a power-law
dependence, as might be expected in the stellar wind from the
progenitor, is discussed in the literature (Bloom 2011).
The GRB afterglow emission sensitively depends on the
circumburst density (Ciardi \& Loeb 2000; Gou et al. 2004; Inoue et
al. 2007), and this is one of the crucial uncertainties in making
predictions for Pop~III bursts.
We next derive the likely circumburst
conditions in the two possible Pop~III host systems.

\subsection{Minihalo case}

For simplicity, we assume that Pop~III GRBs are triggered in minihalos
close to the cooling threshold for collapse, which is only weekly
dependent on redshift (Yoshida et al. 2003). For definiteness, we
choose $M_{\rm vir} \simeq 10^{6}M_{\odot}$. To determine the
pre-burst density, we need to consider the build-up of the \ion{H}{2}
region around the central Pop~III star, and how the density structure
is modified through the strong photoionization-heating inside of it
(Whalen et al. 2004).
The production of ionizing photons
strongly depends on the stellar mass, which in turn is determined by
how the accretion flow onto the growing protostar proceeds under the
influence of this radiation field (e.g., McKee \& Tan 2008; Hosokawa et al. 2011; Smith et
al. 2011; Stacy et al. 2012). Thus, the assembly of Pop~III stars
and the development of an \ion{H}{2} region around them
proceed simultaneously and affect each other. The shallow potential
wells in the host minihalos, with corresponding circular velocities
of a few km\,s$^{-1}$, are unable to retain photo-ionized gas,
so that the gas is effectively blown out of the minihalo. The
resulting photo-evaporation has been studied with 3D radiative transfer
calculations (Alvarez et al. 2006; Abel et al. 2007; Greif et al. 2009),
where one massive Pop~III star at the center of the minihalo acts as
an embedded point source, that also take into account the
hydrodynamic response of the photo-heated gas.

It is possible to
understand the key physics of the photoevaporation from minihalos
with the self-similar solution for a champagne flow (Shu et al.
2002).
Assuming a $\rho \propto r^{-2}$ density profile, which describes
the typical situation in minihalos outside a nearly flat inner core,
one can reformulate the spherically symmetric continuity and Euler
equations for isothermal gas as follows:
\begin{equation} [ (v - x)^2 - 1]{1 \over \alpha} {d \alpha \over d
x}  =
 \left[\alpha - {2 \over x}(x - v) \right](x-v),
\label{alpha} \end{equation}
\begin{equation} [ (v - x)^2 - 1]{d v
\over d x} =
 \left[(x- v)\alpha - {2 \over x} \right](x-v),
\label{vel} \end{equation}
where $x = r /c_s t$ is the similarity variable, and $\rho(r,t) =
\alpha(x)/4\pi G t^2=m_{\rm H} n(r)/X$ and
$u(r,t)=c_s v(x)$ introduce the reduced density and velocity, respectively.
$c_s$ is the sound speed of the
ionized gas and $X=0.75$ the hydrogen mass fraction. The density globally
decreases with time, after the ionizing source has turned on.
To derive the immediate pre-explosion value, we will therefore set
$t=t_{\ast}\simeq 3 \times 10^6$\, yr, the typical lifetime of a massive
Pop~III star (e.g., Bromm et al. 2001b).

Different solutions are obtained
depending on the ratio $\epsilon \equiv
(c_{s,i}/c_s)^2$, where $c_{s,i}$ and $c_s$ are the initial
and ionized isothermal sound speeds. We take $T=3 \times 10^4$ K
and $T_i=T_{\rm vir}$, where the virial temperature is defined as
\begin{equation}
T_{\rm vir}\simeq \frac{GM_{\rm vir} m_{\rm H}}{2R_{\rm vir}k_{\rm B}}\simeq 10^4 {\rm \, K}
\left(\frac{M}{10^8 M_{\odot}}\right)^{2/3}\left(\frac{1+z}{10}\right)\mbox{\ .}
\end{equation}
The intuition behind the latter identification is that $c_{s,i}^2\simeq
k_{\rm B} T_{\rm vir}/m_{\rm H}\simeq v_{\rm vir}^2$,
the specific gravitational potential energy of the gas inside a halo,
is a measure of how strongly the fluid is bottled up, before
a champagne flow can occur.
Using table~1 in Shu et al. (2002), we can obtain the density
profiles at different redshifts as follow. First, the value of
$\epsilon$ can be calculated at the given redshift. We then
derive the corresponding reduced central density, $\alpha_0$, via
interpolation.
For the given
$\alpha_0$ we can numerically solve equations (\ref{alpha}) and
(\ref{vel}) subject to the inner boundary conditions:
\begin{equation}
\alpha = \alpha_0 \quad\mbox{and}\quad v = 0, \quad\mbox{at}\quad x
= 0.
\end{equation}
To avoid the singularity near $x=0$, Shu et al. (2002) provide the series
expansion
\begin{equation}
\alpha=\alpha_0+{\alpha_0\over 6}\left({2\over
3}-\alpha_0\right)x^2+\ldots,
\end{equation}
\begin{equation}
 v={2\over
3}x+{1\over 45}\left({2\over 3}-\alpha_0\right)x^3+\ldots.
\label{seriesexp}
\end{equation}
Once the Pop~III star has turned on at $t=0$, the photo-heating commences,
putting in place a pressurized inner bubble which leads to an outward-flowing
shock at $x_{\rm sh}$. The shock obeys the usual isothermal jump conditions,
written in terms of the reduced variables as
\begin{equation}
(v_u -x_{\rm sh})(v_d - x_{\rm sh)} = 1, \quad {\alpha_d \over \alpha_u} = (v_u -
x_{\rm sh})^2, \label{jump}
\end{equation}
where the subscript $u$($d$) indicates the value of the reduced
velocity and density upstream (downstream) of the shock.

\begin{figure}
\epsscale{1.3}
\plotone{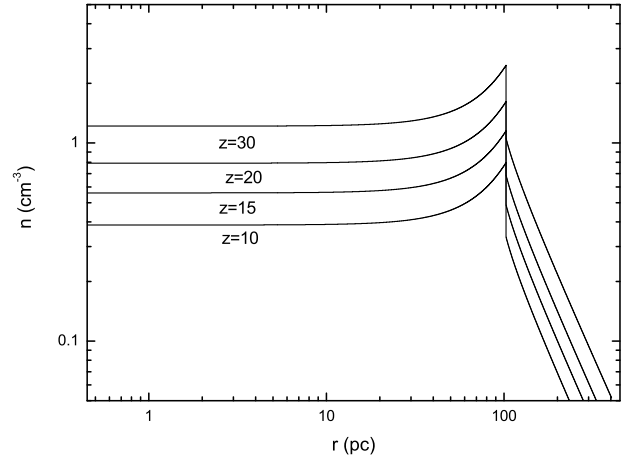} \caption{Minihalo circumburst environment. Shown is the
hydrogen number density vs. distance from the central Pop~III star
at the moment of its death, 3~Myr after it was born. The density
profiles are given by the Shu solution for different redshifts, as labelled.
Typical circumburst densities are $\sim 1$\,cm$^{-3}$, with smaller
values at lower redshifts. The redshift trend reflects the shallower
minihalo potential wells at later times.}
\label{Shu}
\end{figure}

In Figure \ref{Shu}, we show the resulting density profiles at the
end of the Pop~III progenitor's life, and therefore at the completion
of the photoionization-heating as well. As can be seen, the circumstellar
densities have dropped significantly from the high values, $\ga 10^4$\,cm$^{-3}$, present prior to Pop~III star formation,
and are nearly uniform at small radii. Such a flat density profile
is markedly different from that created by stellar winds. The absence of any wind around
a Pop~III star is indeed predicted, as a consequence of very low metallicity (Baraffe et al. 2001;
Kudritzki 2002), and could serve as an indicator of a low-$Z$ GRB progenitor. The post-photoheating
density evolves with redshift, approximately according to
$n_{\rm pi}\propto (1+z)$, normalized such that $n\simeq 1$\,cm$^{-3}$
at $z\simeq 20$. This law differs from earlier estimates of the
Pop~III circumburst evolution, where no $z$-dependence was assumed,
or a scaling with the density of the background universe, $n_{\rm pi}\propto
(1+z)^3$ (e.g., Gou et al. 2004).

This relation can be understood as follows.
Prior to the onset of protostellar
collapse, the baryonic density profile in the minihalo can be described by
the Lane-Emden equation for an isothermal sphere
\begin{equation}
\frac{1}{r^2}\frac{d}{dr}(r^2\frac{d\phi}{dr})=4\pi G\rho_0
e^{-\phi/a^2},
\end{equation}
where $\rho_0$ is the density at which the gravitational potential $\phi$
is set to zero. A simple analytical solution to this equation is the
well-known singular isothermal sphere (SIS):
$\rho_{\rm SIS}=a^2/(2\pi G r^2)$, where $a^2\simeq k_{\rm B}T_{\rm
vir}/m_{\rm H}$. This law approximately describes the profile
found in realistic simulations of first star formation (e.g., Yoshida et
al. 2003), outside an inner core of radius $r_{\rm J}\sim 1$\,pc,
the Jeans length of the primordial gas (Bromm et al. 2002). The baryonic mass
inside the shock radius, $r_{\rm sh}=c_s t_{\ast}\simeq 100$\,pc, is about
\begin{equation}
M_i=\int_{r_J}^{r_{\rm sh}}4\pi r^2 \rho_{\rm SIS}(z,r)dr \mbox{\ .}
\end{equation}
At the end of photoionization-heating, the mass in the flat
inner core is about
\begin{equation}
M_f=\frac{4\pi}{3}r_{\rm sh}^3\rho_{\rm pi}=\frac{4\pi}{3}r_{\rm sh}^3
n_{\rm pi}m_{\rm H}X.
\end{equation}
If we neglect any outflows at large radii, we have approximate mass
conservation, such that
$M_i\simeq M_f$, resulting in $n_{\rm pi}\propto
T_{\rm vir}\propto (1+z)$, as before.

Our post-photoheating densities are consistent with simulation results
(Kitayama et al. 2004; Alvarez et al. 2006; Abel et al. 2007; Greif et al.
2009, 2010). We also agree with Whalen et al. (2004), if we take into
account that their halo had a somewhat lower mass,
$M_{\rm vir}\simeq 5\times 10^5 M_{\sun}$.
Empirical support is provided in Chandra et al. (2010) who found that the
circumburst density of GRB~090423 at $z=8.2$ was about $n\sim 0.9$\,cm$^{-3}$ by
fitting the radio, X-ray and infrared afterglow, although this burst almost
certainly did not originate in a minihalo environment.

\subsection{Atomic cooling halo case}

\begin{figure}
\epsscale{1.3}
\plotone{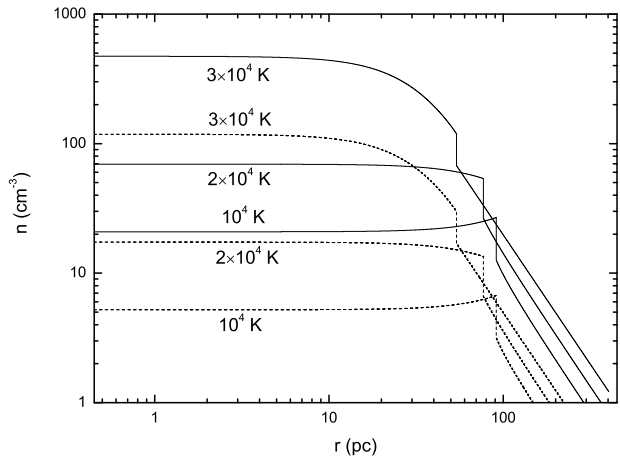} \caption{Atomic-cooling-halo circumburst environment.
The density profiles are again calculated from the Shu solution, evaluated at
$t=3 \times 10^6$\,yr after source turn-on. The collapse redshift
is here indirectly given by the virial temparature, as labelled.
We show the case of photoheating from only a single Pop~III star
({\it solid lines}), and that from a stellar cluster
({\it dotted lines}). The resulting densities again exhibit a nearly
uniform inner core, but overall values are much higher than in the
minihalo case (see Fig.~1).
}\label{atomic}
\end{figure}
Star formation and radiative feedback inside atomic cooling halos is
significantly less well-understood than the minihalo case (e.g.,
Johnson et al. 2009; Safranek-Shrader et al. 2010, 2012), and there are
no comparable high-resolution simulations yet. What is the character of
star formation, in terms of IMF and stellar multiplicity (Clark et al.
2011a)? Given current uncertainty, we here again use the formalism
of the Shu solution as developed above. We further assume for simplicity that
either a single Pop~III star forms, or a small stellar cluster, with
properties that would again heat the surrounding \ion{H}{2} region
to $T\simeq 3\times 10^4$\,K. We stress that the numbers derived here
only provide us with rough guesses at best, and the hope is that
simulations will eventually become available to firm them up.

If a cluster of Pop~III stars forms, the time where photoheating
stops will be prolonged. We thus need to evaluate
$\rho(r,t) = \alpha(x)/4\pi G t^2$ at $t= t_{\rm
cluster}\simeq t_{\ast}+t_{\rm SF}$, where $t_{\rm SF}$ is the
timescale over which star formation is ongoing,
and $t_{\ast}$ again is the lifetime of a single Pop III star.
We can estimate $t_{\rm SF}\sim t_{\ast}$, because disruptive
feedback effects will tend to terminate star formation when the first
member stars die. The resulting densities at the end of photoheating
are shown in Figure~\ref{atomic}. Similar to the minihalo case, densities
are nearly constant at small radii, but overall values are much higher.
The reason is that here the potential wells are deeper, so that
photoheated gas can more easily be retained.
Typical circumburst densities are $n_{\rm pi}\sim 100$\,cm$^{-3}$.
Because such large densities would correspond to high afterglow fluxes,
Pop~III GRBs originating in atomic cooling halos may be extremely
bright, rendering them visible out to very high redshifts.
For some of the most distant bursts known, such large circumburst
densities have indeed been inferred. E.g., Gou et al. (2007)
have argued that fitting the GRB~050904 afterglow at $z=6.3$ requires
a circumburst density of a few $100$\,cm$^{-3}$. Similar values
have been suggested for GRB~080913 at
$z=6.7$ (Greiner et al. 2009; Zhang et al.
2009).

\section{Modeling pre-galactic enrichment}
\begin{figure*}
\centering \subfigure[]{
\begin{minipage}[b]{0.9\textwidth}
\includegraphics[width=0.5\textwidth]{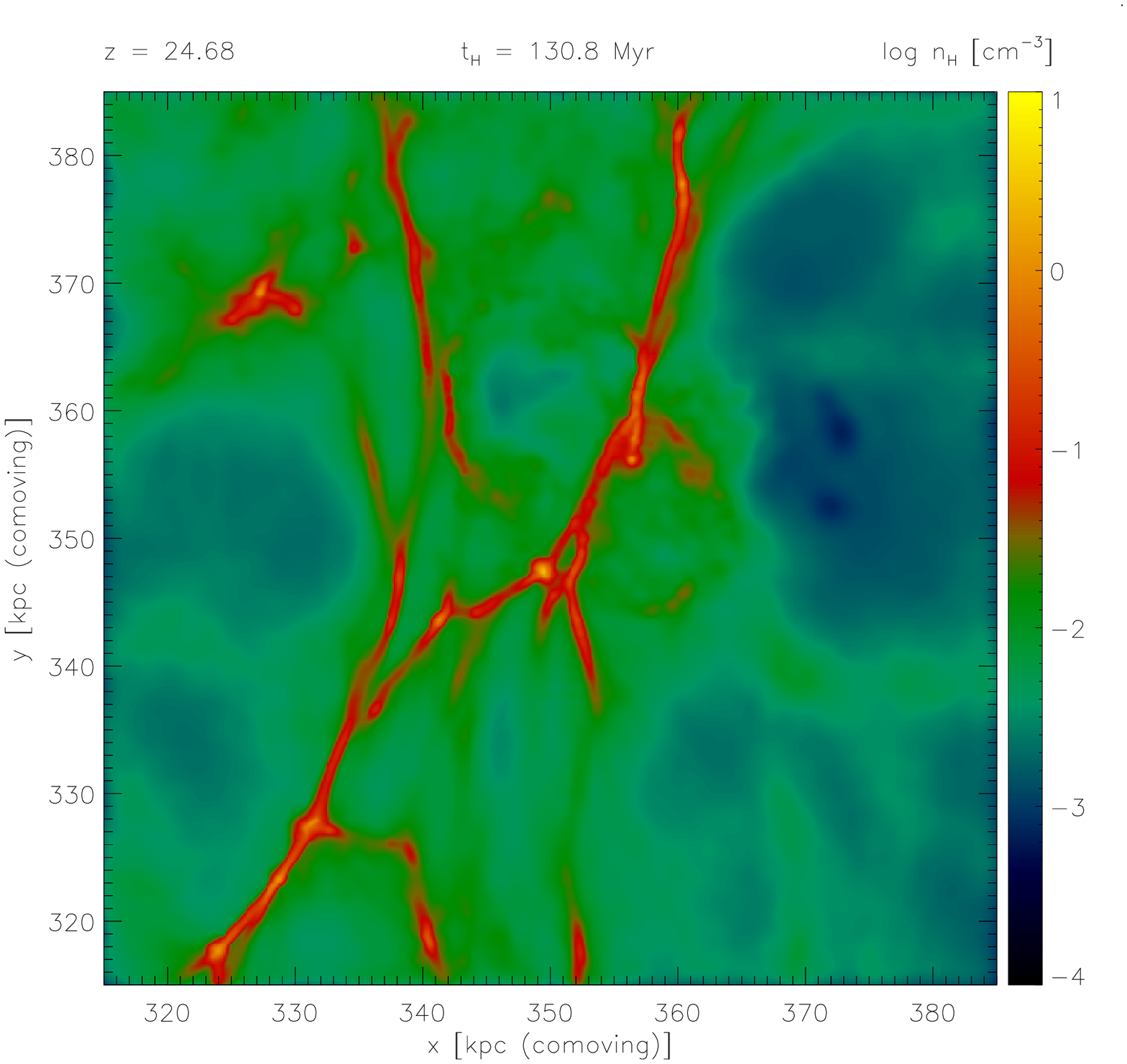}
\includegraphics[width=0.5\textwidth]{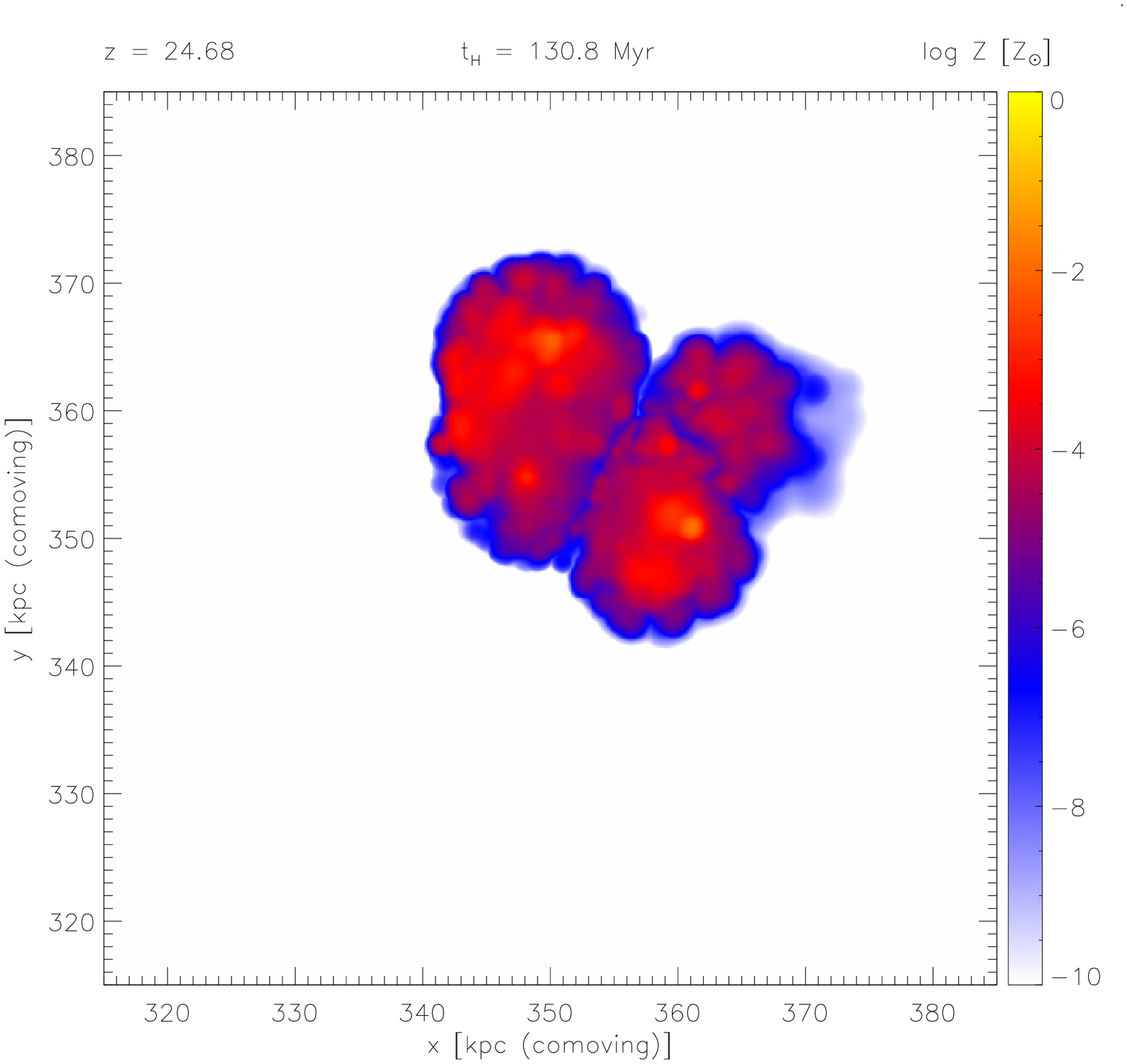}
\end{minipage}
} \subfigure[]{
\begin{minipage}[b]{0.9\textwidth}
\includegraphics[width=0.5\textwidth]{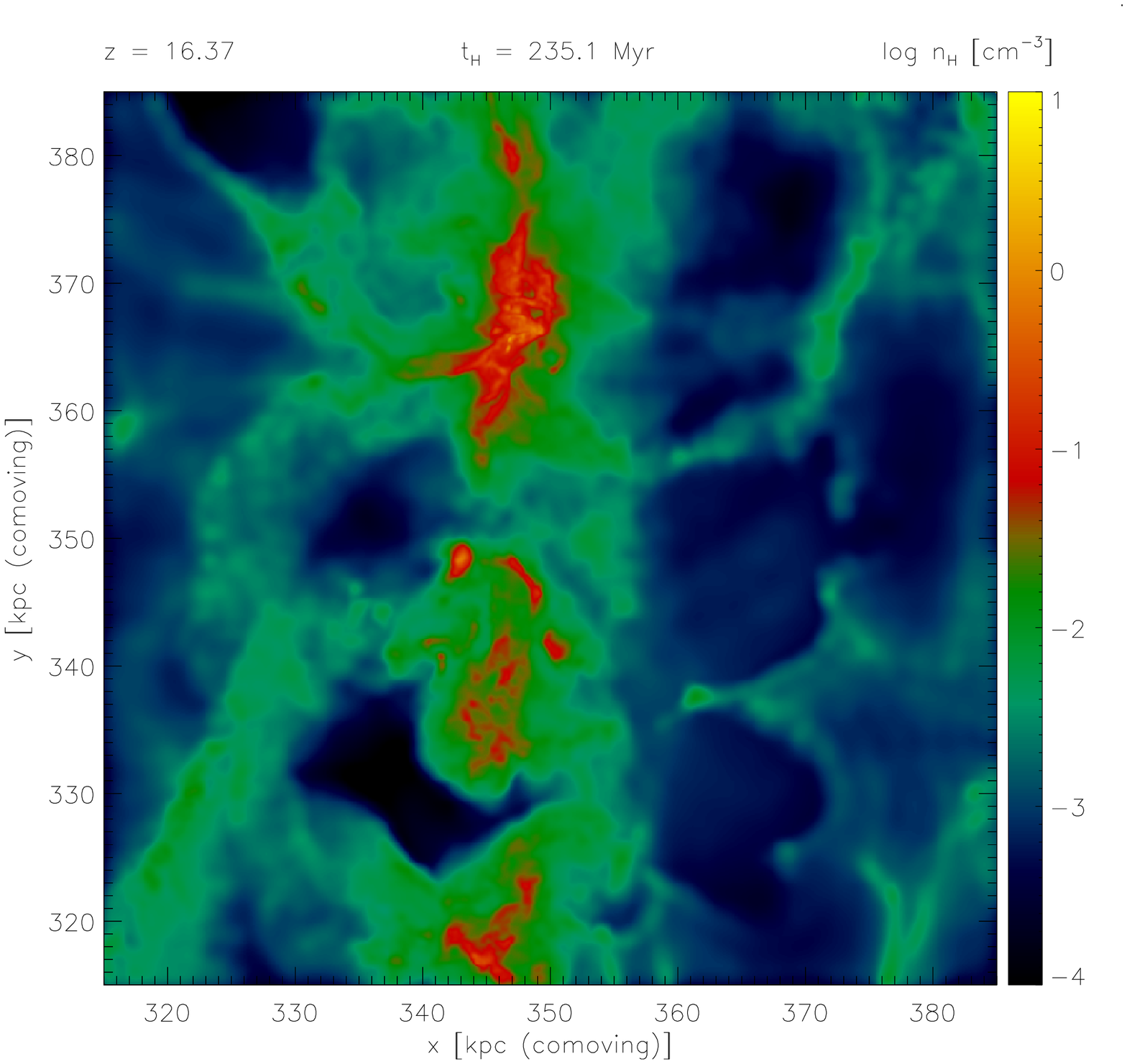}
\includegraphics[width=0.5\textwidth]{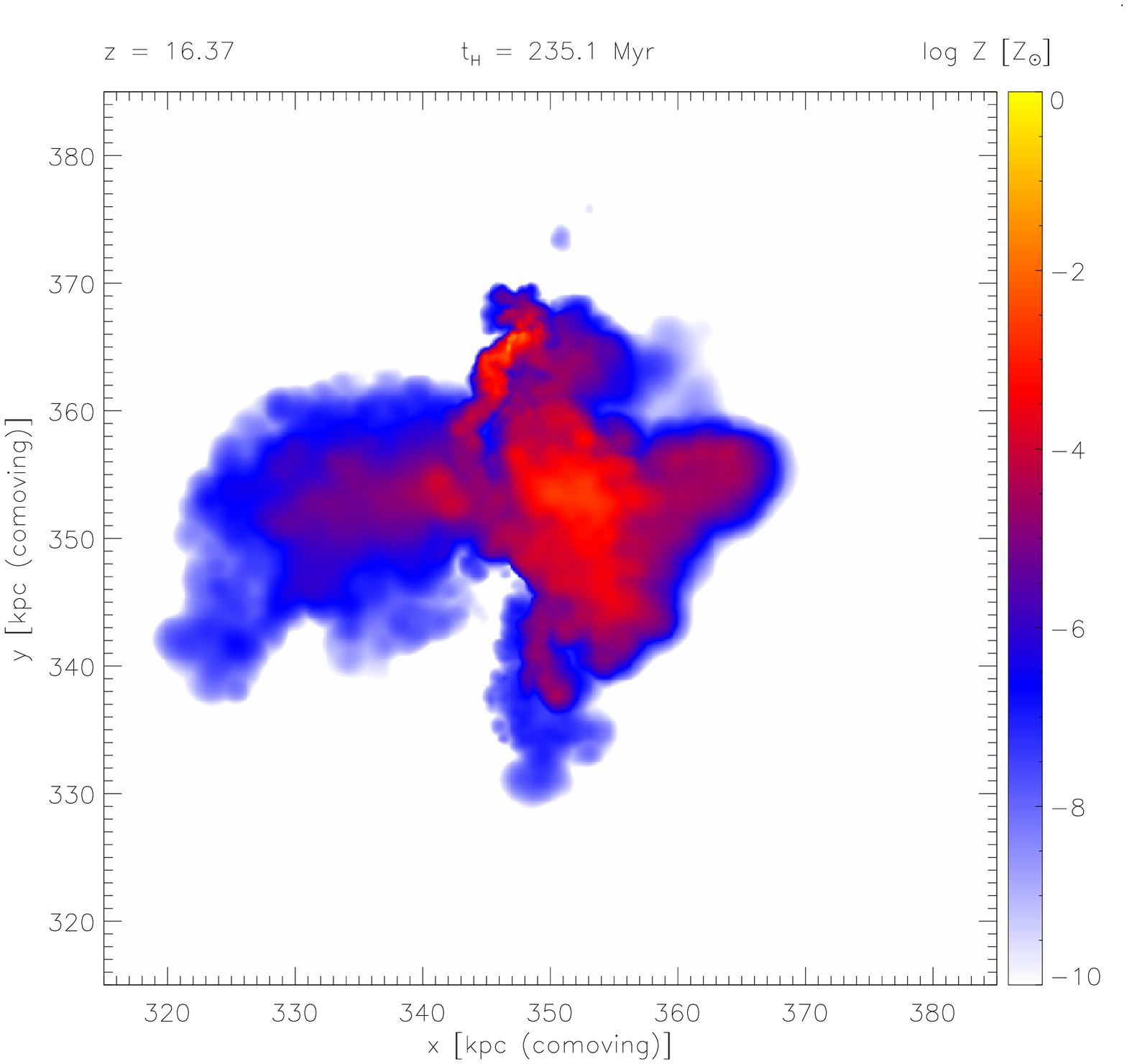}
\end{minipage}
}\caption{Possible explosion sites for high-$z$ GRBs. Shown are the
hydrogen number density and metallicity contours during the assembly
of a first galaxy,
averaged along the line of sight within the central $\simeq 100$\,kpc
(comoving), at two different redshifts. {\it Panel~(a)}: $z\simeq 25$,
briefly after the first Pop~III SN exploded. At this time, most of the
metals reside in the IGM.
{\it Panel~(b)}: $z\simeq 16.4$, closer to the virialization
of the atomic cooling halo. Now, metals are being re-assembled
into the growing potential well of the first galaxy.
The topology of metal enrichment is highly
inhomogeneous, with pockets of highly enriched material embedded in
regions with a largely primordial composition.
These plots are derived from the simulation described in Greif et al.
(2010).
} \label{galsim}
\end{figure*}

In order to predict afterglow spectra from Pop~III GRBs, we consider
explosions that are embedded in a realistic cosmological setting.
Specifically, we use the first galaxy simulation carried out
by Greif et al. (2010), where the assembly of an atomic cooling halo
was tracked, resolving all prior Pop~III star formation in the
progenitor minihalos, together with the radiative feedback from
\ion{H}{2} regions around those stars. In addition, the simulation
allowed one of the Pop~III progenitor stars to explode as an
energetic SN, explicitly following the transport and mixing of
the ejected metals into the IGM. We here summarize a few key aspects
of the simulation, and refer the reader to Greif et al. (2010) for
further details.

\begin{figure}
\epsscale{1.2}
\plotone{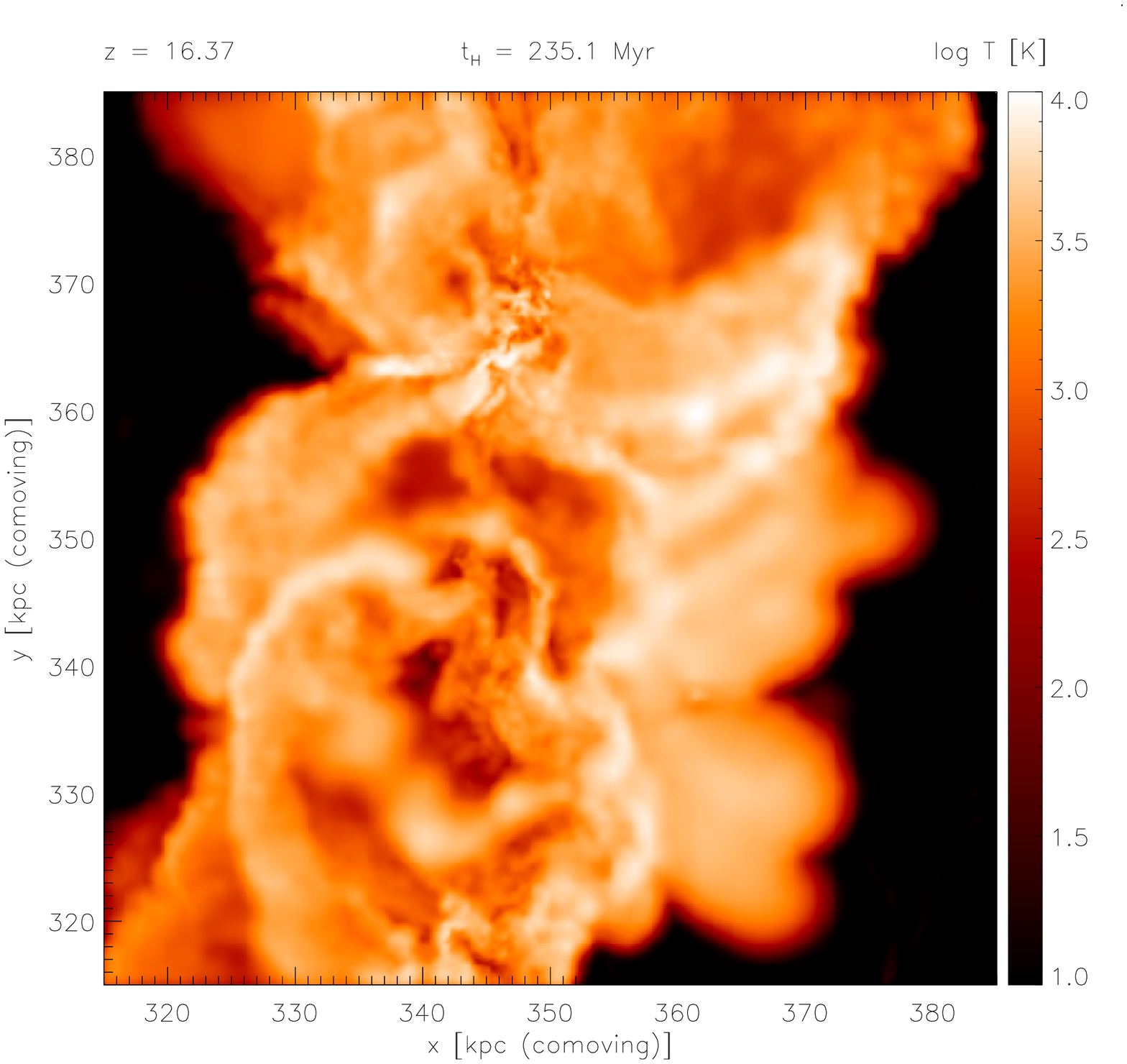} \caption{Environment for high-$z$ GRBs. Shown are temperature contours along the
line of sight within the central $\simeq 100$\,kpc (comoving) at
$z=16.4$, corresponding to panel {\it (b)} in Fig.~\ref{galsim}.
The center of the growing galaxy is highly turbulent, with pockets
of cold gas embedded in a hotter, diffuse medium.
}\label{tem}
\end{figure}
The simulation is carried out in a cosmological box of size
$ 1~\rm{Mpc}$ (comoving), and is initialized at $z=99$
according to a concordance $\Lambda$ cold dark matter ($\Lambda$CDM)
model with matter density
$\Omega_{\rm{m}}=1-\Omega_{\Lambda}=0.3$, baryon density
$\Omega_{\rm{b}}=0.04$, Hubble parameter
$h=H_{0}/\left(100~\rm{km}~\rm{s}^{-1}~\rm{Mpc}^{-1}\right)=0.7$,
spectral index $n_{\rm{s}}=1.0$, and normalization $\sigma_{8}=0.9$
(Spergel et al. 2003). Density and velocity perturbations are
imprinted at recombination with a Gaussian distribution, and they are
propagated to $z=99$, when the simulation is started, by applying the
Zeldovich approximation. Within the highest resolution region
of our nested, zoomed-in initial conditions, the
DM and gas particle masses are $33 M_{\odot}$ and $5 M_{\odot}$, respectively,
The baryonic mass resolution, roughly given by the mass contained inside an SPH
kernel, is $\simeq 400 M_{\odot}$, close to the relevant Jeans mass in the
primordial star forming gas.
To capture the chemical
evolution of the gas, the simulation follows the abundances of H, H$^{+}$,
H$^{-}$, H$_{2}$, H$_{2}^{+}$, He, He$^{+}$, He$^{++}$, and e$^{-}$,
as well as the five deuterium species D, D$^{+}$, D$^{-}$, HD and
HD$^{+}$. All relevant cooling mechanisms are included, i.e., H and He
atomic line cooling, bremsstrahlung, inverse Compton scattering, and
collisional excitation cooling via H$_{2}$ and HD (Johnson \& Bromm
2006).

In Figure~\ref{galsim}, we show the hydrogen number density and
metallicity averaged along the line of sight within the central
$\simeq 100$ kpc (comoving) at two different output times, from
$z\simeq 25$, briefly after the first star-forming minihalo has collapsed,
to $z=16.4$, closer to the virialization of the first galaxy. The
distribution of metals produced by the first SN explosion is
highly inhomogeneous, and the metallicity can reach up to
$Z\sim 10^{-2.5} Z_{\sun}$, which is already larger than any of the
values discussed for the critical metallicity,
$Z_{\rm crit}\la 10^{-4} Z_{\sun}$ (e.g., Bromm et
al. 2001a; Bromm \& Loeb 2003; Frebel et al. 2007), including
those predicted for a dust-driven Pop~III/Pop~II transition, where
$10^{-6} Z_{\sun}< Z_{\rm crit} < 10^{-5} Z_{\sun}$
(Omukai et al. 2005; Schneider et al. 2006). The implication is that
both Pop~III and
Pop~I/II stars will form during the assembly of the first
galaxies (Johnson et al. 2008; Maio et al. 2010), in turn giving rise to
the simultaneous occurrence of Pop~III and normal GRBs at a given
redshift (Bromm \& Loeb 2006; de~Souza et al. 2011). Here, we will focus on the case of a
Pop~III burst exploding in one of the (still metal-free) first galaxy progenitor
minihalos at $z\simeq 16.4$ (see Section~5). In Figure~\ref{tem}, we show
the temperature distribution in the vicinity of the explosion site, allowing
us to calculate (thermal) line broadening in a realistic fashion.

\section{Pop~III Afterglow Emission}

\subsection{Basic Physics}\label{sec:hydro}

Our modeling of the GRB broad-band afterglow emission follows standard
prescriptions, and we here only briefly summarize the basic physics
involved (for general reviews, see
M\'{e}sz\'{a}ros 2006; Zhang 2007).
We specifically consider a relativistic shell ejected from the progenitor of a
GRB (e.g., Kobayashi 2000).
The shell has a rest mass of $M_0$, an observed thickness of
$\Delta_0$, and exhibits an initial Lorentz factor of $\Gamma_0=E_{\rm iso}/M_0 c^2$,
where $E_{\rm iso}$ is the isotropic-equivalent explosion energy.
The interaction between the shell and the surrounding medium with a
density of $n_0$
produces a forward and a reverse shock. The shocks, in turn,
accelerate electrons to high energies, distributed according to a power-law:
$dN(\gamma_e)/d\gamma_e \propto \gamma_e^{-p}$ for
$\gamma_e\geq \gamma_m$, where
$\gamma_e$ is the internal Lorentz factor for the electrons.
The minimum
Lorentz factor is set by assuming that a fraction $\epsilon_e$ of the
post-shock energy is transferred to the electrons, resulting in:
$\gamma_m =
\epsilon_e \gamma (m_p/m_e)(p-2)/(p-1)$, where $\gamma \sim \Gamma_0$ is the Lorentz
factor of the shocked fluid (e.g., Blandford \& McKee 1976; Sari et al. 1998).
If one also assumes that a fraction $\epsilon_B$ of the post-shock
energy is deposited into random magnetic fields, the gyration of the
shock-accelerated electrons around the magnetic field will generate synchrotron radiation, thus
giving rise to the afterglow emission. There is then a second characteristic
Lorentz factor, $\gamma_c$, determining the
threshold below which
electrons do not lose a significant fraction of their energy to radiation.
Equating the synchrotron cooling time with the expansion time of the shell,
one finds:
$\gamma_c m_e c^2=P(\gamma_c)t_{\rm
exp}$, where $P(\gamma_e)=(4/3)\sigma_Tc\gamma_e^2(B^2/8\pi)$
is the synchrotron radiation power (Rybicki \& Lightman 1979), and
$t_{\rm exp}$ the expansion time, measured in the comoving fluid frame (Sari et al.
1998; Panaitescu \& Kumar 2000). Therefore, the cooling Lorentz
factor can be written as: $\gamma_c=6\pi m_e c/(\sigma_T B^2 t_{\rm exp})$.

\begin{figure}
%\epsscale{0.2} \includegraphics{f6N.eps} \caption{Flux density at
\includegraphics[width=0.5\textwidth]{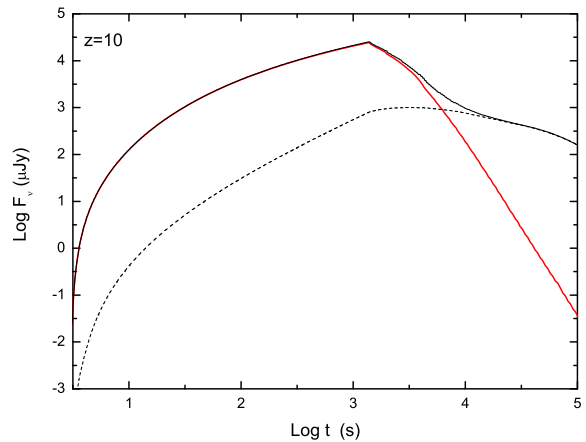} \caption{Flux density at
$\nu=6.3\times 10^{13}$Hz (M band) as a function of observed time.
We show the emission from the forward shock ({\it dashed line}), the
reverse shock ({\it solid red line}), and their combination ({\it
solid black line}). We use the parameters given in the text, for a
GRB exploding in a minihalo with circumburst density
$n_0=0.5$\,cm$^{-3}$ at $z=10$. It can be seen that the reverse
shock dominates during the first few hours, reaching a brightness in
excess of $\sim 1$\,mJy.}\label{LCM}
\end{figure}

For the resulting spectrum, there are now two cases:
In the fast cooling case ($\gamma_m>\gamma_c$), the flux can be
calculated from
$$
F_\nu = F_{\nu,{\rm max}}\left\{
\begin{array}{ll}
(\frac{\nu_a}{\nu_c})^{1/3}(\frac{\nu}{\nu_a})^2 \phantom{-----} \nu<\nu_a \\
(\frac{\nu}{\nu_c})^{1/3} \phantom{-------}\nu_a<\nu<\nu_c \\
(\frac{\nu}{\nu_c})^{-1/2} \phantom{------} \nu_c<\nu<\nu_m \\
(\frac{\nu_m}{\nu_c})^{-1/2}(\frac{\nu}{\nu_m})^{-p/2}
\phantom{--}\nu_m<\nu,
\end{array} \right.
$$

whereas in the opposite, slow-cooling
($\gamma_m<\gamma_c$), case, one has
$$
F_\nu = F_{\nu,{\rm max}}\left\{
\begin{array}{ll}
(\frac{\nu_a}{\nu_m})^{1/3}(\frac{\nu}{\nu_a})^2 \phantom{-----} \nu<\nu_a \\
(\frac{\nu}{\nu_m})^{1/3} \phantom{--------}\nu_a<\nu<\nu_m \\
(\frac{\nu}{\nu_m})^{-(p-1)/2} \phantom{-----} \nu_m<\nu<\nu_c \\
(\frac{\nu_c}{\nu_m})^{-(p-1)/2}(\frac{\nu}{\nu_c})^{-p/2}
\phantom{--}\nu_c<\nu.
\end{array} \right.
$$

The break frequencies correspond to the electron Lorentz factors
introduced above. The frequency $\nu_a$ describes the onset of
synchrotron self-absorption, which is important at low (radio)
frequencies. The afterglow emission exhibits a different character
before and after the reverse shock (RS) has crossed the shell. This
happens at a time $t_\oplus\sim
(1+z)\Delta_0/2c=16.7\mbox{\,s\,}(1+z)\Delta_{0}/10^{12}$\,cm, as measured in the
observer frame. Before RS crossing ($t<t_\oplus$), we have for the
forward shock (``$f$''):
\begin{eqnarray}
\nu_{m,f} &\propto& ~ (1+z)^{-1}  \epsilon_{e}^2
\epsilon_B^{1/2} \Delta_{0}^{-3/2} E_{\rm iso}^{1/2} t^{-1}, \nonumber\\
\nu_{c,f} &\propto& ~ (1+z)^{-1} \epsilon_B^{-3/2}
n_0^{-1} \Delta_{0}^{-1/2} E_{\rm iso}^{-1/2} t^{-1}, \nonumber\\
\nu_{a,f} &\propto& ~ (1+z)^{-1}
\epsilon_B^{6/5} n_0^{11/10} \Delta_{0}^{-1/2} E_{\rm iso}^{7/10} t^{-2}, \nonumber\\
F_{\nu,{\rm max},f} &\propto& ~ (1+z) \epsilon_B^{1/2} n_0^{1/2} E_{\rm iso}
d_{L}^{-2} t,~
\end{eqnarray}
where
$d_L$ is the luminosity distance. Here and in the following we equate
the pre-shock density with that established by photoionization-heating,
at the time of the Pop~III star's death: $n_0\simeq n_{\rm pi}$ (see Section~2).
For the reverse shock (``$r$''), we
have
\begin{eqnarray}
\nu_{m,r} &\propto& ~ (1+z)^{-1}  \epsilon_{e}^2
\epsilon_B^{1/2} \Gamma_0^2 n_0^{1/2}, \nonumber\\
\nu_{c,r} &\propto& ~ (1+z)^{-1} \epsilon_B^{-3/2}
n_0^{-1} \Delta_{0}^{-1/2} E_{\rm iso}^{-1/2} t^{-1}, \nonumber\\
\nu_{a,r} &\propto& ~ (1+z)^{-1}\epsilon_{e}^{6/13}
\epsilon_B^{1/5} n_0^{1/5} \Gamma_0^{-8/5} \Delta_{0}^{-6/5} E_{\rm iso}^{3/5} t^{-3/5}, \nonumber\\
F_{\nu,{\rm max},r} &\propto& ~ (1+z) \epsilon_B^{1/2} n_0^{1/4}
\Gamma_0^{-1} \Delta_{0}^{-3/4} E_{\rm iso}^{5/4} d_{L}^{-2} t^{1/2}.
\end{eqnarray}
After RS crossing, we similarly have
\begin{eqnarray}
\nu_{m,f} &=& \nu_{
m,f}(t_{\oplus})\left(\frac{t}{t_{\oplus}}\right)^{-3/2}, \nonumber\\
\nu_{c,f} &=& \nu_{
c,f}(t_{\oplus})\left(\frac{t}{t_{\oplus}}\right)^{-1/2}, \nonumber\\
\nu_{a,f} &=&  \nu_{a,f}(t_{\oplus})\left(\frac{t}{t_{\oplus}}\right)^{0}, \nonumber\\
F_{\nu,{\rm max},f} &=& F_{\nu,{\rm
max},f}(t_{\oplus})\left(\frac{t}{t_{\oplus}}\right)^{0},
\end{eqnarray}
for the forward shock. For the reverse shock, we have
\begin{eqnarray}
\nu_{m,r} &=&  \nu_{
m,r}(t_{\oplus}) \left(\frac{t}{t_{\oplus}}\right)^{-3/2}, \nonumber\\
\nu_{c,r} &=&  \nu_{
c,r}(t_{\oplus})\left(\frac{t}{t_{\oplus}}\right)^{-3/2}, \nonumber\\
\nu_{a,r} &=&  \nu_{a,r}(t_{\oplus})\left(\frac{t}{t_{\oplus}}\right)^{-1/2}, \nonumber\\
F_{\nu,{\rm max},r} &=& F_{\nu,{\rm
max},r}(t_{\oplus})\left(\frac{t}{t_{\oplus}}\right)^{-1}.
\end{eqnarray}
The exact expressions, and further details,
can be found in Kobayashi (2000) and Wu et
al. (2003). Note that the latter study assumes the presence of a
circumburst wind profile, whereas we here consider the constant-density
(ISM) case, similar to Kobayashi (2000).
The basic parameters of the afterglow emission region are:
$\Gamma_0=300$, $E_{\rm iso}=10^{53}~\mbox{erg}$, $
\Delta_0=10^{12}~\mbox{cm}$, $\epsilon_e=0.3$, $\epsilon_B=0.1$, and
$p=2.5$. We have verified that our afterglow predictions are
consistent with those in Gou et al. (2004), and Inoue et al. (2007), as
these papers focus on Pop~III bursts at high redshifts as well. As an
example, in Figure~\ref{LCM}, we show the M-band light curve, corresponding
to $\nu=6.3\times 10^{13}$\,Hz. It can be seen that the reverse shock
emission dominates during the first few hours, whereas the forward shock
component takes over at later times.

\subsection{Redshift Dependence}

\begin{figure}
\epsscale{1.2}
\plotone{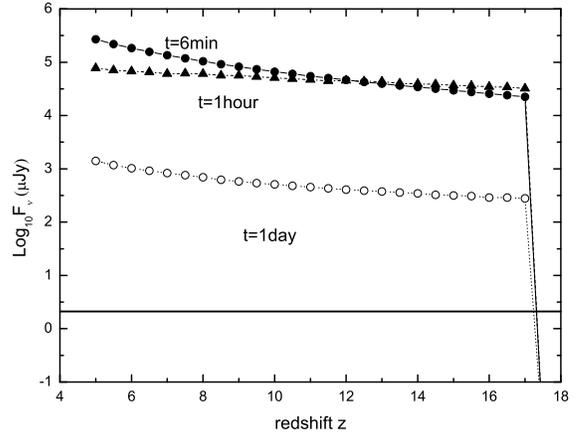} \caption{Observed flux at $\nu=1.36\times
10^{14}$Hz (K band) as a function of redshift.
The lines indicate
the flux at different observed times, as labelled. Here, we
have assumed that the Pop~III GRB is triggered in a minihalo
environment, as modelled in Section~2.1.
We mark the K-band
sensitivity of the NIRSpec instrument on board the {\it JWST}
with a horizontal line. The sharp
cut-off at $z\simeq 17$ is due to Ly$\alpha$
absorption in the IGM.}\label{LCK}
\end{figure}
\begin{figure}
\epsscale{1.2} \plotone{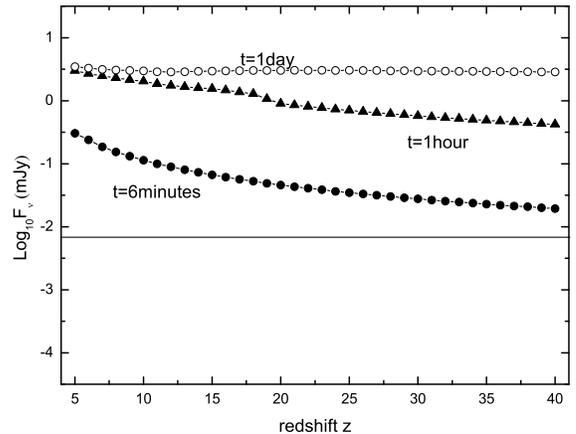} \caption{Observed radio flux (in
mJy) at $\nu=5$\,GHz as a function of redshift, again for a minihalo
burst environment. We employ the same convention for the lines and
symbols as in Fig.~\ref{LCK}, but we now mark the corresponding
sensitivity ({\it horizontal line}) for the EVLA. As is evident, the
radio flux increases with observed time, and is well within reach of
detection.} \label{radio}
\end{figure}

We here briefly summarize the key aspects of afterglows triggered by
Pop~III stars. Figure~\ref{LCK} shows the observed flux at
$\nu=1.36\times10^{14}$\,Hz as a function of redshift
for Pop III GRBs in a minihalo. According to Section~2.1,
circumburst densities can then be written as
$n\simeq 1 \mbox{\,cm}^{-3} (1+z)/20$.
The lines with filled dots, black triangles and open dots
correspond to an observed time of $6$ minutes, 1 hour and 1 day
respectively. The straight line marks the K-band sensitivity for the
near-infrared spectrograph (NIRSpec) on \emph{JWST}, estimated
for a resolution of $R=1000$, a signal-to-noise ratio of S/N$=10$, and
an integration time of 1 hour (Gardner et al. 2006). Once the
observed frequency corresponds to an emitted frequency above the
Ly$\alpha$ resonance,
$\nu_{\alpha}=2.47\times10^{15}$\,Hz, which occurs at $z\simeq 17$
for the K-band, all flux will be
completely absorbed by the intervening, still largely neutral IGM.
As can be seen, in the $K$ band the {\it JWST} will be able to
detect GRBs, and to conduct spectroscopy on their afterglows, out to
$z\sim 16$ even after 1 day. In the $M$ band, the redshift horizon
is extended further still, to $z\sim 35$.
At these frequencies, the
afterglow flux is only weakly dependent on redshift. The main
reason is that cosmic time dilation implies that a given observed time
after the trigger corresponds to successively earlier emission times,
where intrinsic afterglow luminosities are much brighter
(Ciardi \& Loeb 2000; Bromm \& Loeb 2012). Secondly,
circumburst densities modestly increase with redshift (see Section~2).

In Figure~\ref{radio}, we show the flux at $\nu=5$\,GHz, again after
$t=6$ minutes, 1 hour and 1 day. We calculate the
sensitivity of the VLA, using the following
expression (Ioka \& M\'{e}sz\'{a}ros 2005)
\begin{eqnarray}
F_{\nu}^{\rm sen}&=&\frac{{\rm S/N} \cdot 2 k_{\rm B} T_{\rm sys}} {A_{\rm
eff} \sqrt{2 t_{\rm int} \Delta \nu}}
\nonumber\\
&\sim& 23 {\rm \mu Jy} \left(\frac{\rm S/N}{5}\right)
\left(\frac{t_{\rm int}}{1 {\rm
day}}\right)^{-1/2} \left(\frac{\Delta
\nu}{50 {\rm \,MHz}}\right)^{-1/2}\mbox{\,,}
\end{eqnarray}
where $t_{\rm int}$ is the integration
time, $\Delta \nu$ the bandwith, and $A_{\rm
eff}/T_{\rm sys}\sim2\times10^6$\,cm$^2$\,K$^{-1}$.
The bandwidth of the Expanded VLA (EVLA)\footnote{http://www.aoc.nrao.edu/evla/}
can be up to 8\,GHz. For a 2\,h
integration in the 8\,GHz band, the EVLA can thus reach a sensitivity of up to
6.4 $\mu$Jy, which is shown as a horizontal line in Figure~\ref{radio}. We
can see that the radio afterglows are easily within reach for the EVLA,
up to $z\sim 40$ (e.g., Chandra et al. 2010).

\subsubsection{Peak fluxes and wavelengths}

\begin{figure}
\epsscale{1.2}
\plotone{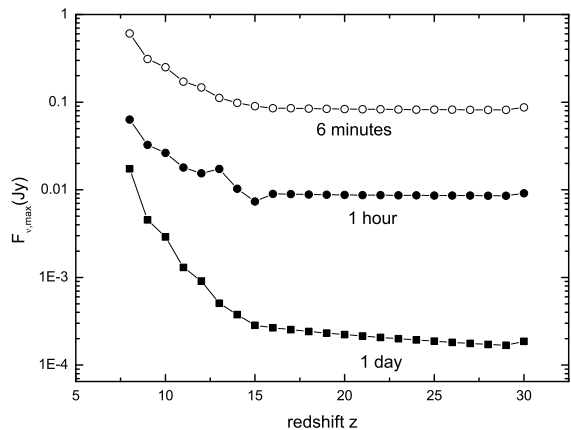} \caption{The observed maximum flux density as a
function of redshift at different times, $t=1$\,day, 1 hour and 6
minutes from bottom to top. We assume that the Pop~III GRB has exploded
in a minihalo. As can be seen, afterglow fluxes can reach
substantial levels early on, and still remain at $\sim$ mJy a day after the trigger, out to very high redshifts.}\label{fluxz}
\end{figure}

\begin{figure}
\epsscale{1.2}
\plotone{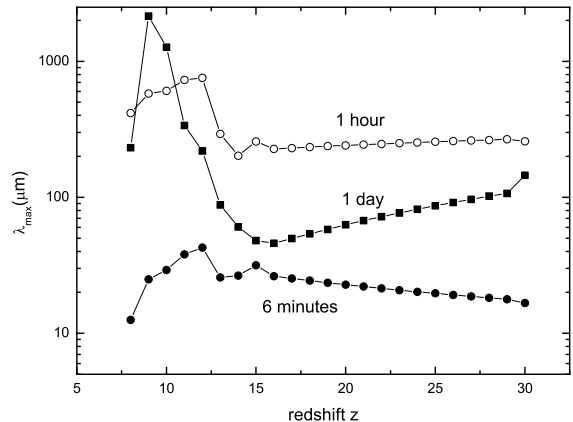} \caption{The observed maximum wavelength as a
function of redshift at different times, as labelled, for a GRB
exploding in a minihalo. The emission maxima range from mm to
mid-IR wavelengths.}
\label{lambdaz}
\end{figure}

According to hierarchical cosmic structure formation,
the characteristic halo mass varies as a function of redshift. Thus,
the typical observed
peak flux and wavelength of Pop~III GRBs are also
a function of redshift. In the following, we derive
ballpark estimates
for these key quantities by considering
$2\sigma$ overdensities (e.g., figure~4 in Clarke \& Bromm 2003).
This provides us with the typical halo mass, $M_{\rm vir}$, at a given
redshift.
In the $2\sigma$ case, halo masses become less
than $10^5 M_{\odot}$ at redshift $z>15$. Since this has dropped below the
threshold mass for H$_2$ cooling (e.g., Yoshida et al. 2003),
Pop~III stars could not form in such low-mass halos.
For higher-sigma peaks, however, super-critical halos
still exist at earlier times. Therefore, for simplicity we
adopt $M_{\rm vir}=10^5 M_{\odot}$ at $z>15$. After the characteristic halo
mass is thus determined, the virial temperature is fixed as well, and the ratio $\epsilon
\equiv (c_{s,i}/c_s)^2$ is known. The Shu solution then gives
the circumburst densities at a given redshift (see Section~2), and
GRB afterglow spectra can finally be calculated.
Figure~\ref{fluxz}
shows the observed peak flux at observed time $t=6$
minutes, 1 hour and 1 day at $z<30$. The peak flux varies
from milli-Jansky to sub-Jansky. Figure \ref{lambdaz} gives the
observational peak wavelength at the same observed times for the same
redshift range. The observed peak wavelengths vary from
millimeter to mid-infrared bands. Early afterglows are best observed
with facilities such as the {\it JWST}, or possibly adaptive-optics
supported next-generation, extremely-large telescopes on the ground. At late
times, radio facilities provide the optimal chance for detection. Among them
is ALMA, which will allow to probe the afterglow in the mm and sub-mm regime, where
the peak of the emission is located about 1~day after the explosion for $z\sim 10$ sources.

\section{Metal Absorption Lines}

As we have seen above, Pop~III GRBs can in principle be detected out
to very high redshifts. Their afterglow spectra provide us with a
unique probe into the state of the early IGM, including its degree
of ionization (e.g., Totani et al. 2006) and metal enrichment. The
latter will be encoded in absorption lines imprinted on the smooth
afterglow emission with a signature that depends on the nature of
the Pop~III SNe. Here, we will consider for simplicity that prior to
the GRB only one nearby SN exploded beforehand, dispersing its
complement of heavy elements into the otherwise pristine IGM. Such a
scenario is consistent with the clustered nature of Pop~III star
formation, as is expected for the high-$\sigma$ peak host systems
(e.g., Greif \& Bromm 2006). Specifically, we here explore
nucleosynthetic yields for Type~II core-collapse SNe,
and for pair-instability supernovae (PISNe).
In Table~1, we summarize the average yields per SN, $Y_{\rm X}$, for
select elements that are expected to dominate the spectral
signature. For Type~II SNe, we use the yields of Woosley \& Weaver
(1995). These depend on the explosion energy, which introduces
uncertainties of $\sim 25\%$. For the PISN scenario, we utilize the
yields of Heger \& Woosley (2002; 2010). Table~1 also lists the
properties of several key transitions, including their rest-frame
wavelength, $\lambda_i$, and their oscillator strength, $f_{{\rm
osc},i}$.

We employ a simple model for the ionization structure of the metal-enriched
gas. If the hydrogen in this region is substantially neutral, metals will
reside in states typical of \ion{H}{1} regions in the Milky Way (\ion{C}{2},
\ion{O}{1}, \ion{Si}{2}, and \ion{Fe}{2} for the elements of
interest), because photons able to further ionize these elements
will be absorbed by \ion{H}{1} (Furlanetto \& Loeb 2003). The
ionization potential of \ion{O}{1} is $13.62$ eV, and it remains
locked in charge exchange equilibrium with \ion{H}{1} (Oh 2002).
A modest extragalactic radiation background at frequencies below the
Lyman limit could easily maintain these metals in a singly-ionized state
(e.g., Furlanetto \& Loeb 2003). We thus assume that C, O, Si, and Fe
are predominantly in the ionization states typical for Galactic \ion{H}{1}
clouds.

\subsection{Cosmological Radiative Transfer}

Employing comoving coordinates, the equation
of cosmological radiative transfer
is (e.g., Abel et al. 1999):
\begin{equation}
\frac{1}{c} \frac{\partial I_{\nu}}{\partial t} + \frac{\hat{n}
\cdot \nabla I_{\nu}}{\bar{a}} - \frac{H(t)}{c} (\nu \frac{\partial
I_{\nu}} {\partial \nu} - 3 I_{\nu}) = j_{\nu} - k_{\nu}
I_{\nu}\mbox{\, ,} \label{equ:crt}
\end{equation}
where $I_{\nu}$ is the intensity
of the radiation field, $\hat{n}$ a unit vector along the
direction of the ray, $H(t) \equiv \dot{a}/a$ the
Hubble constant\footnote{We here employ the same cosmological parameters as in the underlying simulation (see Sec.~3).}, and $\bar{a} \equiv a(t+dt)/a(t)$ the ratio
of cosmic scale factors separated by $dt$.
Here $j_{\nu}$ and $k_{\nu}$
denote the emission and absorption coefficient,
respectively. Equation~(\ref{equ:crt}) can be simplified if
the effects of cosmic expansion are negligible. This will be the case
if the light-crossing time, $t_{\rm light}=L/c$, over the region of interest,
here the (physical) size of our simulation box, $L=\mbox{\,1\,Mpc}/(1+z)$,
is small compared to the Hubble time, $t_{\rm H}\simeq
5\times 10^{8}$\,yr\,$[(1+z)/10]^{-3/2}$. Since $t_{\rm light}\ll t_{\rm H}$,
the transfer equation reverts to its non-cosmological form:
\begin{equation}
\frac{1}{c} \frac{\partial I_{\nu}}{\partial t} + \hat{n} \cdot
\nabla I_{\nu} = j_{\nu} - k_{\nu} I_{\nu} \mbox{\ .}\label{equ:trt}
\end{equation}
\noindent We further neglect any diffuse (re-)emission of radiation, such
that $j_{\nu}=0$, and assume near steady-state conditions ($\partial/
\partial t\simeq 0$),
resulting in the static transfer equation:
\begin{equation}
  \label{eq:srt}
  \hat{n} \cdot\nabla I_{\nu} = -k_{\nu} I_{\nu}.
\end{equation}
\noindent If $I_{\nu}(0)$ is the intensity of the central
afterglow, approximately treated as a point source, we have the standard
solution: $I_{\nu}=I_{\nu}(0) \exp(-\tau_{\nu})$, where $\tau_{\nu}$ is
the optical depth along the line-of-sight (LOS).

We can obtain the optical depth as follows.
For a (physical) differential distance element, $dl$, along the given LOS,
we have: $d\tau_\nu=\sigma_\nu n_{\rm X} dl$, where
\begin{equation}
\sigma_\nu=\sqrt{\pi} e^2/ (m_e c)f_{{\rm osc},i} H(u,x)/\Delta
\nu_{\rm D}
\end{equation}
is the frequency dependent absorption cross-section, and
$n_{\rm X}$ the number density of species X
(e.g., \ion{O}{1}, \ion{C}{2}, \ion{Fe}{2}, \ion{Si}{2},
\ion{H}{1}).
In the above equation,
$\Delta \nu_{\rm D} \equiv b \nu_i/c$ is the Doppler width\footnote{We here neglect any broadening from turbulent motions. This assumption is good in the minihalo progenitors, but breaks down in the emerging atomic cooling halo.} with
parameter $b = \sqrt{2k_{\rm B}T / m_{\rm H}}$.
The Voigt function, $H$, is given by
\begin{equation} \label{eq:vhf}
H(u,x) = \frac{u}{\pi} \int_{-\infty}^{+\infty} \,
        \frac{\mathrm{e}^{-y^2}}{(x-y)^2 + u^2} dy \, .
\end{equation}
Here, the relative strength of natural (damping) to Doppler broadening is
described by the parameter
\begin{equation}
u = \frac{\Gamma_i}{ 4 \pi  \Delta \nu_{\rm D}} \mbox{\ ,}
\end{equation}
where $\Gamma_i$ is the
damping constant for transition $i$ (Morton \& Smith 1973).
The variables $x = (\nu - \nu_i)/{\Delta
\nu_{\rm D}}$ and $y = v/b$ are the frequency difference
relative to the line center in Doppler units, and the
normalized particle velocity, with respect to the GRB location, respectively.
The total LOS value of the optical depth is then:
$\tau_\nu=\int\sigma_\nu n_{\rm X} dl$.
We note that we can neglect any wavelength differences between the GRB source
and the metal absorber, because of the very small expansion-generated redshift
on the scale of our simulation box. Specifically, we estimate that
$\Delta z=\Delta
t/(7.5\times10^{7}$\,yr)$\sim 0.005$, where we assume $z\sim 10$, and
$\Delta t\sim t_{\rm light}\sim 3\times 10^{5}$\,yr.

We extract the optical depths along random lines of sight with the ray-tracing
algorithm of Greif et al. (2009).
We choose $N_\theta=160$ and
$N_\phi=320$ in the spherical coordinates, which corresponds to
51200 rays. For each ray, we use 200 logarithmically-spaced radial bins, and
approximate the optical depth integral with a sum along the LOS.
The Pop~III GRB will likely explode in a still substantially neutral IGM.
Any flux shortward of the Ly$\alpha$ resonance is therefore completely absorbed,
and because of the corresponding very large optical depth, there will be absorption
in the red damping wing as well. We model this with the
analytical formula given by Miralda-Escud\'e (1998), using
$z_{\rm reion}=7$ and standard cosmological parameters
(Komatsu et al. 2011). The line shape is not very sensitive to the
choice of $z_{\rm reion}$ (Bromm et al. 2001b).

\begin{deluxetable*}{cccccc}
  \tablecaption{Supernova Yields and Important Transitions.}
  \tablehead{ \colhead{Element} & \colhead{$Y_{\rm X}$ (Scalo) \tablenotemark{a}}
    & \colhead{$Y_{\rm X}$ (VMS) \tablenotemark{b}} &
    \colhead{Ionization State} & \colhead{$\lambda_i$ (\AA) \tablenotemark{c}} &
    \colhead{$f_{{\rm osc},i}$ \tablenotemark{d}} }
  \tablewidth{14.5cm}
  \startdata
  C & 0.1 M$_\sun$ & 4.1 M$_\sun$ & \ion{C}{2} & 1334.5 & 0.1278 \\
  & & & \ion{C}{4} & 1548.2 & 0.1908 \\
  & & & \ion{C}{4} & 1550.8 & 0.09522 \\
  O & 0.5 & 44 & \ion{O}{1} & 1302.2 & 0.04887 \\
  Si & 0.06 & 16 & \ion{Si}{2} & 1304.4 & 0.094 \\
  & & & \ion{Si}{4} & 1393.8 & 0.514 \\
  & & & \ion{Si}{4} & 1402.8 & 0.2553 \\
  Fe & 0.07 & 6.4 & \ion{Fe}{2} & 1608.5 & 0.058 \\
  & & & \ion{Fe}{2} & 2344.2 & 0.114 \\
  & & & \ion{Fe}{2} & 2382.8 & 0.300 \\
  \enddata
  \tablenotetext{a} {Metal yield in Scalo IMF case.}
  \tablenotetext{b} {Metal yield in VMS IMF case.}
  \tablenotetext{c} {Rest-frame transition wavelength.}
  \tablenotetext{d} {Oscillator strength.}
\end{deluxetable*}

%--------------------------------------------------------------------------------
\begin{table*}
 \caption{Line Flux Ratios from minimum to maximum.} \centering
\begin{tabular}{r|cc|cc}
\hline \hline
observed time & crossing time & crossing time &   1 day & 1 day \\
IMF & VMS & Scalo & VMS & Scalo\\
\hline
\ion{O}{1}/\ion{C}{2} 1334.5\AA & 0.078-0.086 & 0.918-0.965 & 0.079-0.090 & 0.091-0.097\\
\ion{O}{1}/\ion{Fe}{2} 2334.2\AA & 0.014-0.030 & 0.408-0.444 & 0.014-0.042 & 0.040-0.043\\
\ion{O}{1}/\ion{Fe}{2} 2382.8\AA & 0.019-0.031 & 0.424-0.458 & 0.018-0.045 & 0.041-0.045\\
\ion{O}{1}/\ion{Si}{2} 1304.4\AA & 0.045-0.059 & 0.807-0.881 & 0.045-0.089 & 0.081-0.088\\
\ion{O}{1}/\ion{Fe}{2} 1608.5\AA & 0.078-0.090 & 0.800-0.852 & 0.078-0.105 & 0.080-0.085\\
\hline \hline \end{tabular}
\end{table*}
%--------------------------------------------------------------------------------

\subsection{Synthetic Spectra}

\begin{figure}
\includegraphics[width=0.5\textwidth]{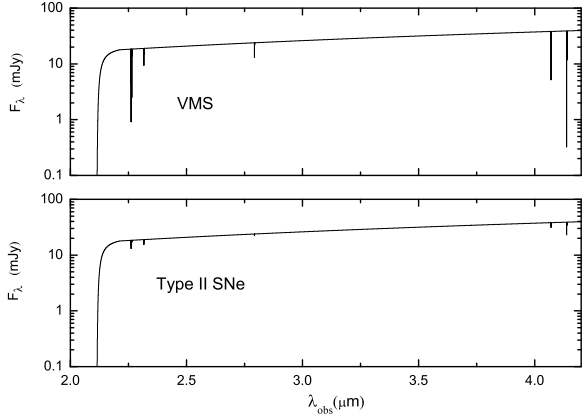} \caption{Total GRB spectrum observed
at the reverse shock crossing time:
$t_{\oplus}=16.7\times(1+16.4)$\,s. We show the spectral region redshifted into the
near IR, accessible to the NIRCam and NIRSpec instruments on board the {\it JWST}.
Metal absorption lines are imprinted according to the Pop~III SN event, PISN vs. core-collapse
(Type~II). The former originates from a very massive star (VMS) progenitor, whereas the latter from a
less massive one. In each case, the cutoff at short wavelengths is due to Lyman-$\alpha$ scattering
in the neutral IGM.
}\label{totalspeTC}
\end{figure}

\begin{figure}
\includegraphics[width=0.5\textwidth]{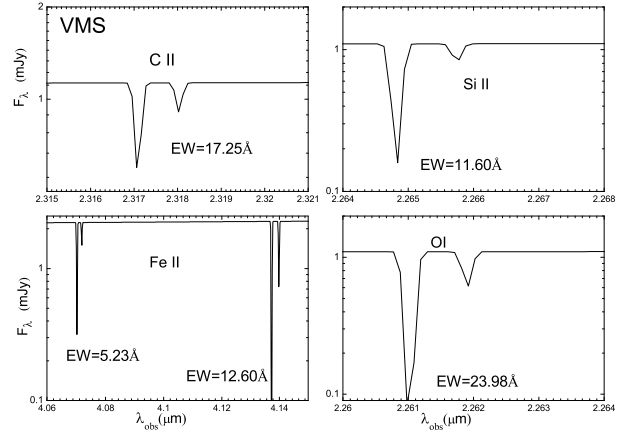} \caption{Spectral signature
of metal
absorption lines at the reverse
shock crossing time: $t_{\oplus}=16.7\times(1+16.4)$\,s. We use the
metal yields in the VMS case. The observed equivalent widths (EWs)
are also shown in the figure. If multiple lines are present, the
given EWs belong to the strongest lines in each panel.
}\label{ewVMSTC}
\end{figure}

\begin{figure}
\includegraphics[width=0.5\textwidth]{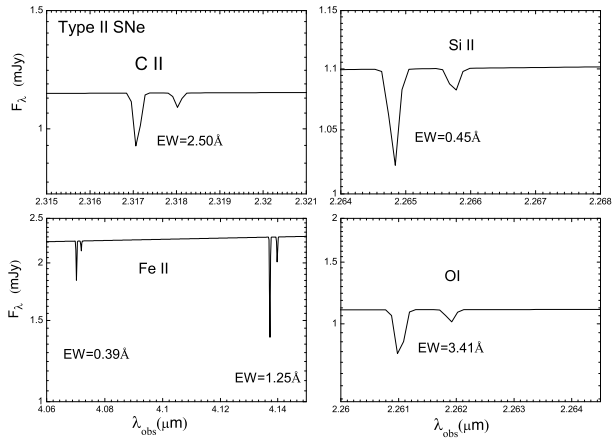} \caption{The same as
Fig.~\ref{ewVMSTC}, but displaying metal absorption lines for a normal (Scalo) Pop~III IMF.
Individual EWs are again given for the strongest lines in each panel.}\label{ewSNeTC}
\end{figure}

In Figure~\ref{totalspeTC}, we show the part of the resulting afterglow spectrum
that has been redshifted into the near-IR, where the {\it JWST} is most sensitive.
For the two assumptions regarding the Pop~III IMF, top-heavy (VMS) and more normal
(Scalo), the spectra are depicted at
the reverse shock crossing time. The metal absorption lines are imprinted on the
smooth underlying GRB afterglow spectrum, rendering their identification straightforward.
The cutoff at short wavelenghts is due to Lyman-$\alpha$ scattering in the IGM
which is expected to be still completely neutral at $z>10$. For detailed predictions,
it is advantageous to quantify the strength of absorption lines through equivalent widths (EWs).
The observed EWs can
be calculated from
\begin{equation}
W=(1+z)\int[1-e^{-\tau(\lambda)}]d\lambda \mbox{\ .}
\end{equation}

In Figures~\ref{ewVMSTC} and \ref{ewSNeTC}, we zoom in on
the vicinity of the strongest lines, again at the reverse
shock crossing time for both IMF cases. In addition, we
explicitly give the derived EWs, which are about
a few ten {\AA} for a top-heavy IMF, and an order of
magnitude smaller for a normal (Scalo) IMF.
The EW values do not change significantly, if evaluated at 1\,day after the
explosion. Such line strengths are well within reach of the
NIRSpec instrument on board the {\it JWST}, which has a spectral
resolution up to $R\equiv \lambda/\Delta
\lambda=1000$, and a line sensitivity of $10^{-18}$\,erg\,s$^{-1}$\,cm$^{-2}$ (Gardner et al. 2006).
In general, the limiting equivalent width of an unresolved line
at observed wavelength $\lambda_{\rm obs}$, at $5\sigma$ significance,
is (Tumlinson et al. 2002):
\begin{equation}
W_{\rm min}=\frac{5\lambda_{\rm obs}}{R\mbox{(S/N)}}\mbox{\ ,}
\end{equation}
where S/N is the signal-to-noise ratio per resolution element.
Within our assumptions, therefore, spectroscopy
of Pop~III GRB afterglows will be able to distinguish between
different enrichment events, PISN vs. core-collapse. There is,
however, a possible complication.
The total enrichment level would be rather similar to that from one PISN for
the case of 10 Type~II SNe. Such a situation is not implausible, given
the number of Pop~III star formation sites, i.e., minihalos, in the
vicinity of the Pop~III GRB (see Sec.~3).
However, we could then still discriminate between
the cases using absorption line flux ratios, which remain quite different.

\begin{figure}
\includegraphics[width=0.5\textwidth]{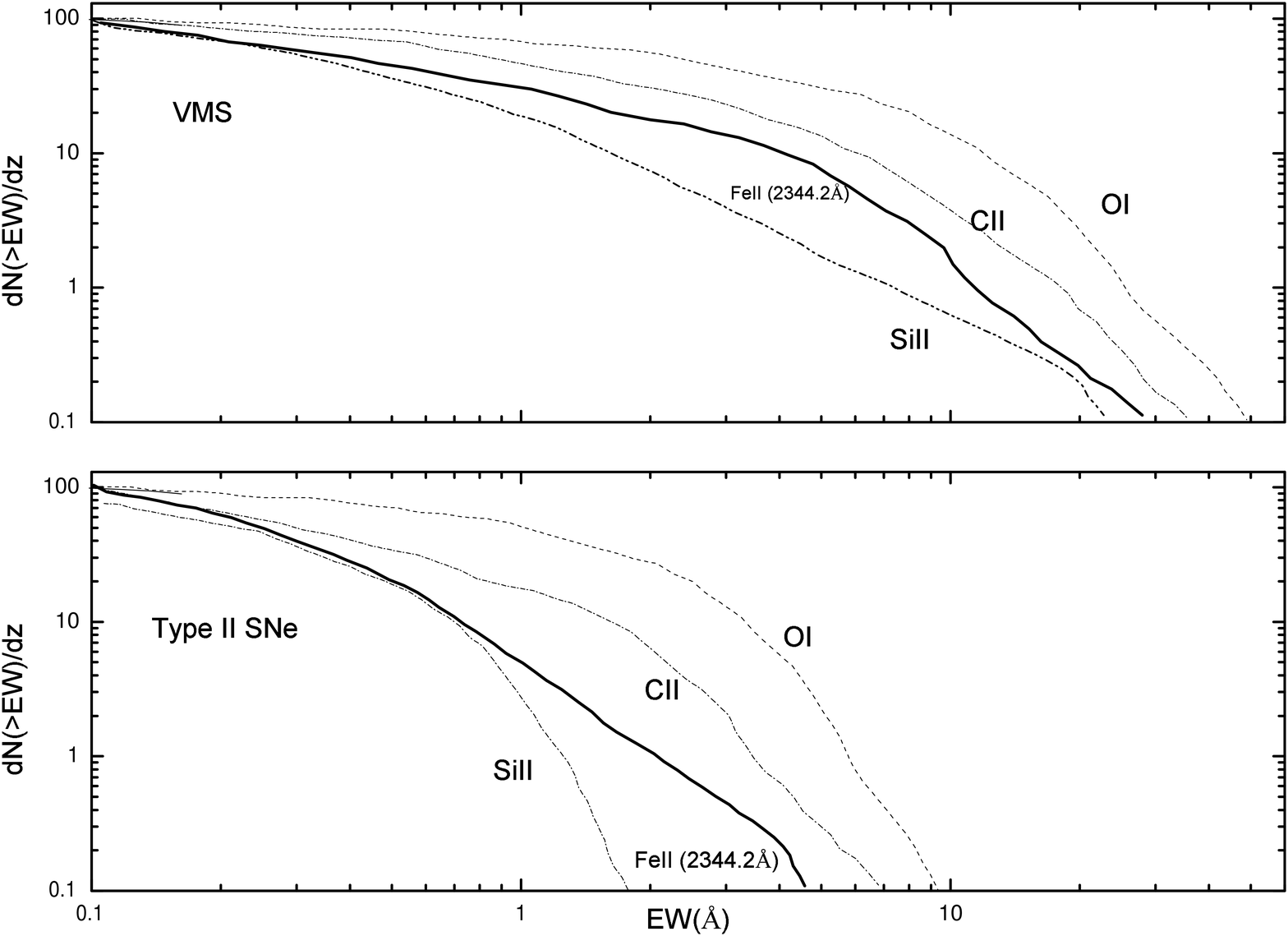}
\caption{Cumulative distribution of metal line equivalent
widths. The curves are calculated by tracing 100 randomly selected
sightlines through the simulation box, centered on the location of the GRB.
{\it Top panel:} VMS case. The curves show the main lines, as labelled.
{\it Bottom panel:} Normal IMF case, corresponding to conventional core-collapse (Type~II) SN
explosions. It is evident that the resulting line statistics is quite distinct for the
two IMF cases.}
\label{linesta}
\end{figure}

In Table~2, we list ranges for select flux ratios, as encountered along different LOSs.
We consider the ratios
\ion{O}{1}/X, where X denotes another species, at both the reverse shock
crossing time and at 1~day after the explosion. In the former case, there
is an order of magnitude difference between the two IMFs, allowing a robust
identification. At later times, however, this clear-cut signature disappears
again.
In Figure~\ref{linesta}, we show the cumulative EW distribution for
100 randomly selected sightlines at redshift $z=16.4$. Again, the
resulting distributions are quite different for the two IMF cases.
It is, however, important to note that the statistical significance
of this analysis is limited by the fact that it is based on
only one simulation, and only one GRB afterglow. In reality, observations
will sample multiple explosion sites with respective LOS distributions that
sample many clouds at various evolutionary stages along the way. Our
analysis should still provide a representative view of the typical
situation, whereas a more complete study has to await the further
increase in available computational power.

\section{Summary and Conclusions}
In this paper, we develop a diagnostic to study pre-galactic metal enrichment in the
vicinity of the first galaxies. Specifically, we utilize
the bright afterglow of a Pop~III GRB
as a featureless background source, and calculate the
strength of metal absorption lines that are imprinted by the first
heavy elements produced by Pop~III SNe. To approximately derive the metal
absorption line statistics, we use an existing highly-resolved
simulation of the formation of a first galaxy which is characterized
by the onset of atomic hydrogen cooling in a halo with virial
temperature $\ga 10^4$\,K.
The reverse shock initially dominates the afterglow flux a few
hours after explosion, followed by the forward shock emission later on.
The fluxes in the near-IR and radio bands may be detectable with the
\emph{JWST} and the VLA out to $z>30$.
We predict that the afterglow emission peaks
from near-IR to millimeter bands with peak fluxes from
mJy to Jy at different observed times.
Metal absorption lines in the GRB afterglow spectrum, giving rise to
EWs of a few tens of {\AA},
may allow us to
distinguish whether the first heavy elements were produced in a Pop~III star that died as
a PISN, or a core-collapse SN. To this extent, the spectrum needs to
be obtained sufficiently early, within the first few hours after
the trigger.

The absorption signature of the first SN events might allow us to constrain
the underlying mass scale of Pop~III stars. This is the
key input parameter to predict their evolution, nucleosynthetic yields,
and modes of death, which in turn determine how the first stars impact
subsequent cosmic history. The strength of the associated stellar feedback
governs the assembly process of the first galaxies, in the sense that
the stronger feedback from more massive stars shifts their formation
to later stages in the hierarchical build-up of structure (e.g., Ricotti et
al. 2002; Greif et al. 2010; Frebel \& Bromm 2012; Ritter et al. 2012; Wise et al. 2012).

It will, however, be very challenging to directly
probe the initial epoch of cosmic star formation. The reason is that even
the {\it JWST} will not be able to detect individual first stars, but instead
only more massive stellar systems or clusters that form later on in more
massive host systems (e.g., Pawlik et al. 2011). Pop~III GRBs may thus
afford us one of the few direct windows into the crucial epoch of first light,
another one being the extremely energetic pair-instability SN or hypernova
explosions that are predicted to end the lives of the most massive Pop~III
stars (e.g., Pan et al. 2012; Hummel et al. 2012). The promise provided by
the combination of a wide field GRB trigger mission, such
as JANUS or Lobster, with the next-generation of highly-sensitive near-IR telescopes,
such as the {\it JWST} or the planned ground-based extremely large facilities (the E-ELT\footnote{http://www.eso.org/sci/facilities/eelt/}, GMT\footnote{http://http://www.gmto.org/}, and TMT\footnote{http://http://www.tmt.org/}), is huge. GRBs are likely to play a key
role in finally opening up the high-redshift frontier, all the way back to the very beginning
of star and black hole formation.

\acknowledgements We are indebted to Anna Frebel and Steve
Finkelstein for helpful discussions. This work is supported by the
National Natural Science Foundation of China (grants 11103007 and
11033002). V.~B. acknowledges support from NSF grant AST-1009928 and
NASA ATFP grant NNX09AJ33G. K.~S.~C. is supported by the GRF Grants
of the Government of the Hong Kong SAR under HKU 7011/09p. The
simulations presented here were carried out at the Texas Advanced
Computing Center (TACC).

\end{document}